\documentclass[aps,prl,reprint,groupedaddress,showpacs,floatfix]{revtex4-1}
\usepackage{graphicx}% Include figure files
\usepackage{dcolumn}% Align table columns on decimal point
\usepackage{bm}% bold math
\begin{document}

\title{Formation and Stability of Cellular Carbon Foam Structures:\\
       An {\em Ab Initio} Study}

% Please use the appropriate macro for each type of information


\author{Zhen Zhu}
\affiliation{Physics and Astronomy Department,
             Michigan State University,
             East Lansing, Michigan 48824, USA}

\author{David Tom\'anek}
\email[E-mail: ]{tomanek@pa.msu.edu}
% \homepage[home page:]{http://www.pa.msu.edu/~tomanek/}
\affiliation{Physics and Astronomy Department,
             Michigan State University,
             East Lansing, Michigan 48824, USA}

\date{\today}
%\begin{linenumbers}

%---------------------------------------------------------------------
\begin{abstract}
We use {\em ab initio} density functional calculations to study
the formation and structural as well as thermal stability of
cellular foam-like carbon nanostructures. These systems with a
mixed $sp^2$/$sp^3$ bonding character may be viewed as bundles of
carbon nanotubes fused to a rigid contiguous 3D honeycomb
structure that can be compressed more easily by reducing the
symmetry of the honeycombs. The foam may accommodate the same type
of defects as graphene, and its surface may be be stabilized by
terminating caps. We postulate that the foam may form under
non-equilibrium conditions near grain boundaries of a
carbon-saturated metal surface.
\end{abstract}
%---------------------------------------------------------------------


% \keywords{boron, density functional theory, electronic structure}


\maketitle
The last few decades have witnessed an unprecedented interest in
carbon nanostructures, the most prominent of them being
fullerenes\cite{Kroto85}, nanotubes\cite{Iijima91} and
graphene\cite{Geim04}. Previously postulated hybrid carbon
nanofoam
structures\cite{DT141,Seifert2006,Zhao2011,Ivanovskii2011} with a
mixed $sp^2$/$sp^3$ bonding character have received much less
attention for lack of direct experimental observation. The growing
body of information about the formation of carbon nanostructures
including graphene\cite{Banhart2011},
nanotubes\cite{Banhart2007,LinNL2007} and fibers\cite{Helveg2004}
on transition metal surfaces with a particular morphology suggests
ways that should favor the formation of particular carbon
allotropes. We propose that previously unseen nanostructures
including carbon foam may form under specific conditions on a
metal substrate.

Inspired by previously postulated carbon
foams\cite{DT141,Seifert2006,Zhao2011,Ivanovskii2011}, we explore
ways to grow such structures on a carbon saturated metal
substrate. We use {\em ab initio} density functional calculations
to investigate the equilibrium structure, structural and thermal
stability and elastic properties of the growing system. The foam
structures we study, which have a mixed $sp^2$/$sp^3$ bonding
character and resemble a bundle of carbon nanotubes fused to a
contiguous 3D honeycomb structure, are rather stable even as slabs
of finite thickness. The foam structure may be compressed more
easily by reducing the symmetry of the honeycombs. It may
accommodate the same type of defects as graphene at little energy
cost, and its surface may be stabilized by terminating caps. We
postulate that the foam could form under non-equilibrium
conditions near grain boundaries of a carbon-saturated metal
surface and should remain stable until $T{\agt}3,500$~K.

%=============================
% \section{Computational Method}
%=============================

Our calculations of the equilibrium structure, stability, elastic
properties and the formation mechanism of the carbon foam were
performed using {\em ab initio} density functional theory (DFT) as
implemented in the \textsc{SIESTA} code\cite{SIESTA}. We used the
Ceperley-Alder \cite{Ceperley1980} exchange-correlation functional
as parameterized by Perdew and Zunger\cite{Perdew81},
norm-conserving Troullier-Martins
pseudopotentials\cite{Troullier91}, and a double-$\zeta$ basis
including polarization orbitals. We used periodic boundary
conditions for the 3D infinite foam structure and 2D slabs of
finite thickness. The 3D foam with 10 atoms per unit cell was
sampled by a fine grid\cite{Monkhorst-Pack76} of at least
$16{\times}16{\times}16$~$k$-points in the Brillouin zone. We used
a mesh cutoff energy of $180$~Ry to determine the self-consistent
charge density, which provided us with a precision in total energy
of ${\alt}2$~meV/atom.

%===========< FIGURE 1 >=================================================
% entire page. There is no need to do explicit centering.
\begin{figure}[!tb]
\includegraphics[width=1.0\columnwidth]{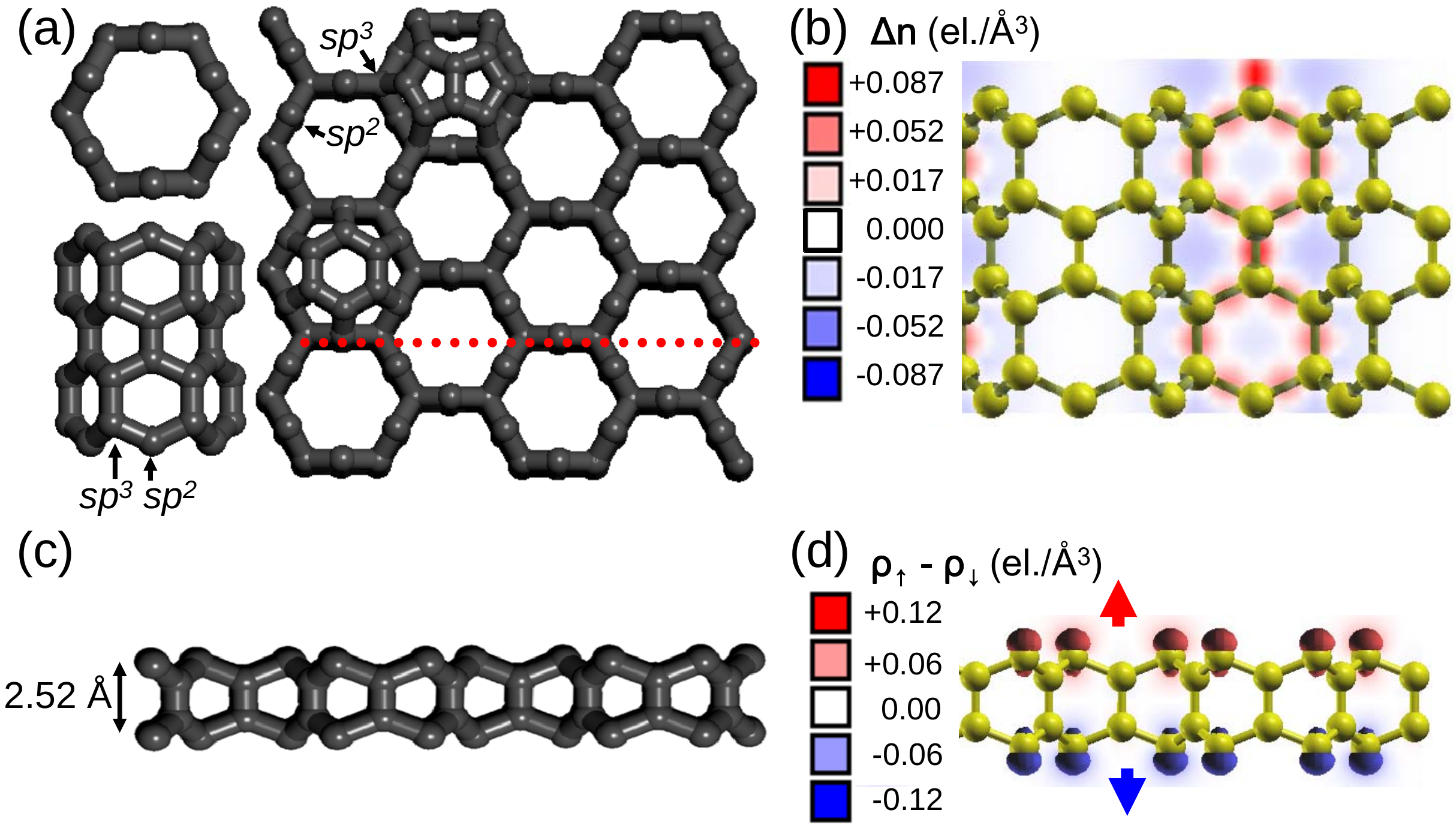}
\caption{(Color online) Structure and electronic properties of
cellular carbon nanofoam. (a) Left panels depict individual cells
of the foam in top and side view. Right panel shows the contiguous
foam in top view, with individual cells terminated by different
types of caps. (b) Electron density difference ${\Delta}n({\bf
r})$ in a plane normal to the surface, indicated by the dotted
line in (a). (c) Side view of the structure of a stable
minimum-thickness foam slab. (d) Spin density distribution
$\rho_\uparrow-\rho_\downarrow$ in the structure shown in (c),
represented in the same plane as in (b). The isosurfaces are shown
for $\rho_\uparrow-\rho_{\downarrow}={\pm}0.05$~el./{\AA}$^3$.
\label{fig1} }
\end{figure}
%===========< FIGURE 1 >=================================================

The structure of the carbon foam is depicted in
Fig.~\ref{fig1}(a). In top view, it closely resembles the graphene
honeycomb lattice with two important distinctions. We find the
optimum lattice constant in the honeycomb plane of the foam to be
$a=4.81$~{\AA}, which is about twice the graphene value
$a=2.46$~{\AA}. More important, 1D carbon-carbon bonds in the 2D
graphene structure correspond to 2D walls in the infinite 3D foam
structure. The foam cells, shown in the left panels of
Fig.~\ref{fig1}(a), closely resemble (6,0) carbon nanotubes. The
foam contains 60\% 3-fold coordinated C atoms, labeled $sp^2$, and
40\% 4-fold coordinated C atoms, labeled $sp^3$. The gravimetric
density of the optimized foam structure, ${\rho}=2.4$~g/cm$^3$,
lies in-between the experimental values\cite{Kittel05} for
graphite, ${\rho}=2.27$~g/cm$^3$, and diamond,
${\rho}=3.54$~g/cm$^3$. We find the 3D carbon foam structure to be
less stable than graphene by
${\Delta}E_{coh}{\approx}0.42$~eV/atom,
% E_coh(bulk foam) = -8.27 eV/atom
% E_coh(graphene) = -8.69 eV/atom
which is comparable to the C$_{60}$ fullerene.

In Fig.~\ref{fig1}(b) we display the electron density difference,
defined by
${\Delta}n({\bf{r}})=n_{tot}({\bf{r}})-{\sum}n_{atom}({\bf r})$ as
the difference between the total electron density
$n_{tot}({\bf{r}})$ and the superposition of atomic charge
densities $n_{atom}({\bf r})$. Charge accumulation in the bond
region indicates strong covalent bonding especially between
neighboring $sp^2$ atoms. Our DFT results for the electronic
structure indicate that the bottom of the conduction band lies
below the top of the valence band, suggesting that the infinite
foam should be metallic. In reality, this finding is a well-known
artifact of DFT that we correct using the LDA+U method, which
indicates semiconducting behavior of the bulk structure.

Besides the bulk structure, we also considered and optimized foam
slabs of different thickness. We must take into account the fact
that the surface terminated with $sp^2$-type atoms, which are
shared by two honeycombs, is inequivalent to a surface with
$sp^3$-type atoms, which are shared by three honeycombs. The
thinnest stable free-standing slab, dubbed the `single-decker'
structure and shown in Fig.~\ref{fig1}(c), has both surfaces of
the $sp^3$-type. It has some commonalities with graphitic
nanostructures that show magnetic ordering at zigzag
edges\cite{{Okada03},{Higuchi04},{AlonsoZZ08},{HodACSNN08},{DT160}}.
Similar to the narrowest zigzag graphene nanoribbon, our system
displays a flat band near $E_F$ that gives rise to spin
polarization with antiferromagnetic coupling across the slab, as
seen in Fig.~\ref{fig1}(d). The dominating role of the surface
reduces the stability of the `single decker structure' by
${\Delta}E_{coh}=0.95$~eV/atom with respect to the bulk carbon
foam. We note an even-odd alternation in the energy as a function
of slab thickness in terms of the number of hexagon rows, since
the slab surfaces may be either identical or different. In any
case, the role of the surface decreases with increasing slab
thickness, and reaches a much smaller value
${\Delta}E_{coh}{\approx}0.46$~eV/atom in the `triple-decker',
shown in Fig.~\ref{fig1}(b), than in the `single-decker'
structure.

The energy penalty due to unsaturated surfaces may be
significantly reduced if the slab is attached to a substrate, or
if the cells are covalently terminated by caps, similar to the
dome termination of carbon nanotubes. We considered either a
hexagon or two adjacent pentagons as candidate caps to terminate
the honeycombs, as seen in the right panel of Fig.~\ref{fig1}(a).
Both caps have 6 twofold coordinated C atoms at the edge that may
form covalent bonds with the surface atoms. Assuming that all
honeycombs on one side are capped and using $A=20.04$~{\AA}$^2$
for the area of each honeycomb, we estimated the surface energy
reduction associated with cap termination to be
${\Delta}E_s=-1.03$~eV/{\AA}$^2$
% $6.65$~eV/foam surface atom
for hexagonal caps and %
% $2.50$~eV/foam surface atom
% Etot(both sides capped single decker,26 atoms)=-4012.72 eV,
% Etot(8-atom 2p cap)=-1232.05 eV,
% Etot(single-decker,10 atoms)=-1538.62 eV
${\Delta}E_s=-0.25$~eV/{\AA}$^2$ %
for the less-stable two-pentagon caps. We need to note that this
stabilization energy contains the termination energy of both the
surface and the individual unsaturated caps, and that these energy
terms can not easily be separated.
% Individual unterminated caps are not very stable:
% Hexagon-cap: 6 atoms, Etot = -915.19 eV or Etot=-152.53 eV/atom
% 2-pentagon cap: 8 atoms, Etot = -1232.05 eV or Etot=-154.01 eV/atom
% Graphene: Binding energy Etot = -155.234 eV/atom

Since epitaxy is an issue when considering the possibility of foam
growth on a metal substrate, we investigated the lateral
compressibility of the foam structure. Our definition is analogous
to the elastic response of a uniform isotropic 3D structure with
volume $V$ to hydrostatic pressure $P=F/A$, given by the force $F$
per area $A$, which is represented by the bulk modulus
$B=-V({\partial}P/{\partial}V)_T$. The elastic deformation of the
area $A$ within a 2D slab structure subject to in-plane
hydrostatic pressure $P_{2D}=F/l$, given by the force per length
$l$, can be represented by an analogous 2D bulk modulus, defined
by $B_{2D}=-A({\partial}P_{2D}/{\partial}A)_T$. Of course, we
expect $B_{2D}$ to be nearly proportional to the slab thickness.
We find this value to be quite useful, since it allows to
determine the critical slab thickness for epitaxial growth on a
particular incommensurate substrate.

%===========< FIGURE 2 >=================================================
% entire page. There is no need to do explicit centering.
\begin{figure}[tb]
\includegraphics[width=1.0\columnwidth]{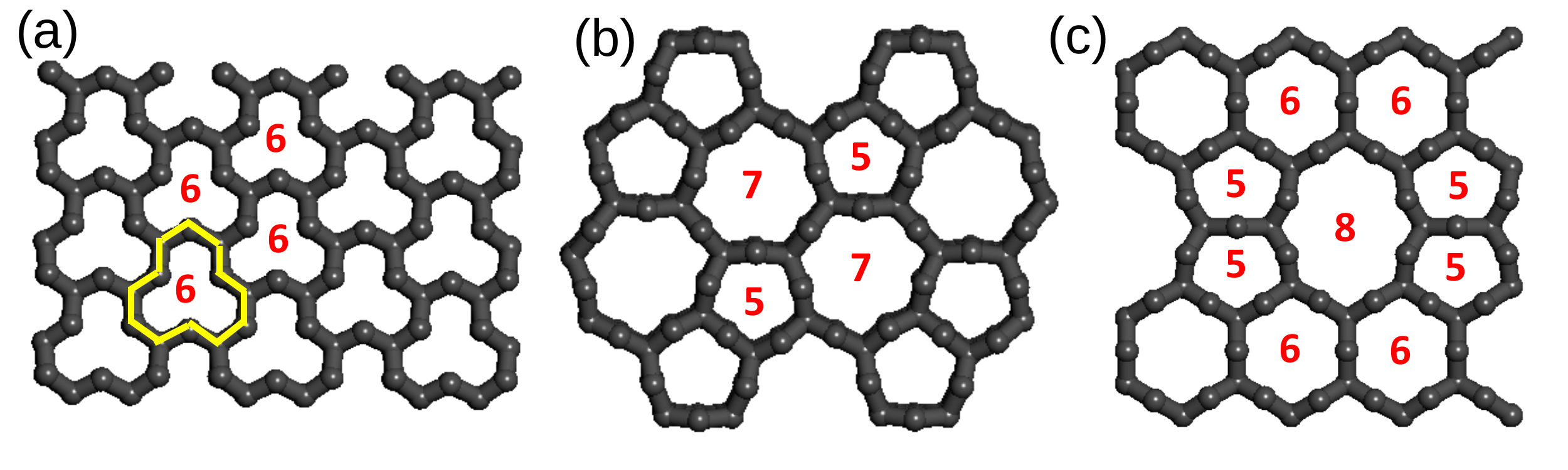}
\caption{(Color online) Defects in the foam. (a) Folding of the
perfect foam, induced by applying hydrostatic pressure or by
electron doping. Foam structures containing (b) 5775 and (c) 558
defects, familiar from defective graphene. \label{fig2}}
\end{figure}
%===========< FIGURE 2 >=================================================

Applying hydrostatic pressure in the plane of the layer, we find
that the honeycomb structure may be compressed more easily by
breaking the honeycomb symmetry than by uniformly compressing the
honeycombs. The structure of the deformed foam, depicted in
Fig.~\ref{fig2}(a), indicates the preferential way the foam may
fold. For this elastic response, we find $B_{2D}=112.9$~N/m in the
`single-decker' and $B_{2D}=163.9$~N/m in the `triple-decker'
structure. For the sake of comparison, when considering a very
thick slab of thickness $h$, we used the bulk calculation to
obtain $B_{3D}{\approx}B_{2D}/h=178$~GPa. We find this value to be
much smaller than that of the ideal structure with suppressed
folding, which had been studied previously\cite{Seifert2006} with
results similar to our value $B_{2D}/h=299.4$~GPa.
% If the pressure were extended up to $200$~Gpa, the folded foam
% would break and form a new structure with low symmetry.
Even though the possibility of folding reduces the bulk modulus,
finite compressibility should still play a significant role during
foam growth on lattice-mismatched or defective substrates.

Interestingly, we find that foam folding occurs spontaneously when
the system is doped by electrons. The structure presented in
Fig.~\ref{fig2}(a) can be obtained by either applying isotropic
pressure in 2D or, at zero pressure, by doping with 0.2 electrons
per C atom. In the latter case, we find that folding induced by
doping reduced the foam energy by $0.19$~eV/atom for the bulk
structure.

We also find that the proposed foam structure may accommodate a
similar type of defects as graphene with the main difference that
bond rotations in graphene correspond to wall rotations in the
foam. In graphene monolayers, lines of 5775 or
Stone-Thrower-Wales\cite{{Stone86},{Thrower69}} and of 558 defects
have been observed to accumulate near grain
boundaries\cite{Huang11,Arkady11,Meyer08} and step
edges\cite{Lahiri10}.
%Lusk2010: http://dx.doi.org/10.1088/1367-2630/12/12/125006
%           extended defects-theory
%Lahiri10: http://dx.doi.org/10.1038/NNANO.2010.53
%           558 defects near step edges on Ni substrate, experimental
%Meyer08: http://dx.doi.org/10.1021/nl801386m
%           5775 defects in free-standing graphene -- TEM
%Arkady11: http://dx.doi.org/10.1021/nn102598m
%           Review of defects
%Huang11: http://dx.doi.org/10.1038/Nature09718
%           CVD grown graphene on Cu has defect lines connecting grains
Their presence reduced stress in strained free-standing layers and
the lattice mismatch energy in adsorbed layers, which in this way
maintained their epitaxy over large areas. The analogous 5775 or
558 defect structures in the foam are depicted in
Fig.~\ref{fig2}(b) and \ref{fig2}(c). Since the foam structure is
rather flexible, the energy penalty associated with these types of
defects is relatively small, amounting to ${\Delta}E=0.19$~eV/atom
for the $5775$ structure of Fig.~\ref{fig2}(b) and
${\Delta}E=0.20$~eV/atom for the $558$ structure of
Fig.~\ref{fig2}(c) with respect to the perfect infinite honeycomb
lattice. With a bulk modulus $B{\approx}250$~GPa, the defective
$5775$ and $558$ foam structures are slightly more compressible
than the perfect foam with suppressed cell folding. Similar to
supported graphene, these types of defects should reduce the
lattice mismatch energy on a particular substrate caused by
different lattice constants or, on a polycrystal, across grains
with different orientation.

To find out whether the carbon foam may or may not decompose to a
more stable allotrope under growth conditions, we studied its
thermal stability by performing molecular dynamics (MD)
simulations in the temperature range
$500$~K${\alt}T{\alt}5,000$~K. To avoid artifacts caused by small
unit cells, we used supercells containing 160 carbon atoms. For
these large unit cells, we used the Tersoff bond-order
potential\cite{tersoff1988} in molecular dynamics simulations
covering time periods of 10~ps using 0.5~fs time steps. Our
results, presented in the EPAPS on-line
material\cite{EPAPS-foamgrowth12}, indicate that the infinite foam
should be stable up to a high melting temperature near $3,700$~K.
Even though free-standing slabs with finite thickness may be
thermally less stable, termination by caps or attachment to a
substrate should increase their thermal stability.

Inspired by the observed growth of graphene\cite{Banhart2011} and
carbon nanotubes\cite{Banhart2007} on cobalt saturated with
carbon, we studied possible growth pathways of the foam on this
substrate. To get insight into the foam-substrate interaction
including optimum lattice registry, we represented the Co(0001)
surface by a four-layer slab with the two bottom layers
constrained in the bulk geometry. Besides the perfect Co(hcp)
lattice, we also considered fcc layer stacking when discussing
grain boundaries. We considered different foam terminations at the
interface in order to find the optimum interface geometry. We
found that the $sp^2$-type terminated foam surface attaches more
strongly to Co(0001) than the $sp^3$-type terminated surface. The
largest reduction of the foam surface energy by
${\Delta}E_s=-0.75$~eV/{\AA}$^2$
% ${\Delta}E_s=-0.745$~eV/{\AA}$^2$ is for an
%          uncapped double-decker on Co
% ${\Delta}E_s=-0.556$~eV/{\AA}$^2$ is for a single-decker,
%          hex-capped on one side and adsorbed on Co on the other side
occurs, when surface C atoms occupy the hollow sites.
% ${\Delta}E_s=-0.83$~eV/{\AA}$^2$for on-top sites
We should note that this stabilization energy reflects the
reduction of both the metal and the foam surface energy.

%===========< FIGURE 3 >=================================================
% entire page. There is no need to do explicit centering.
\begin{figure}[tb]
\includegraphics[width=1.0\columnwidth]{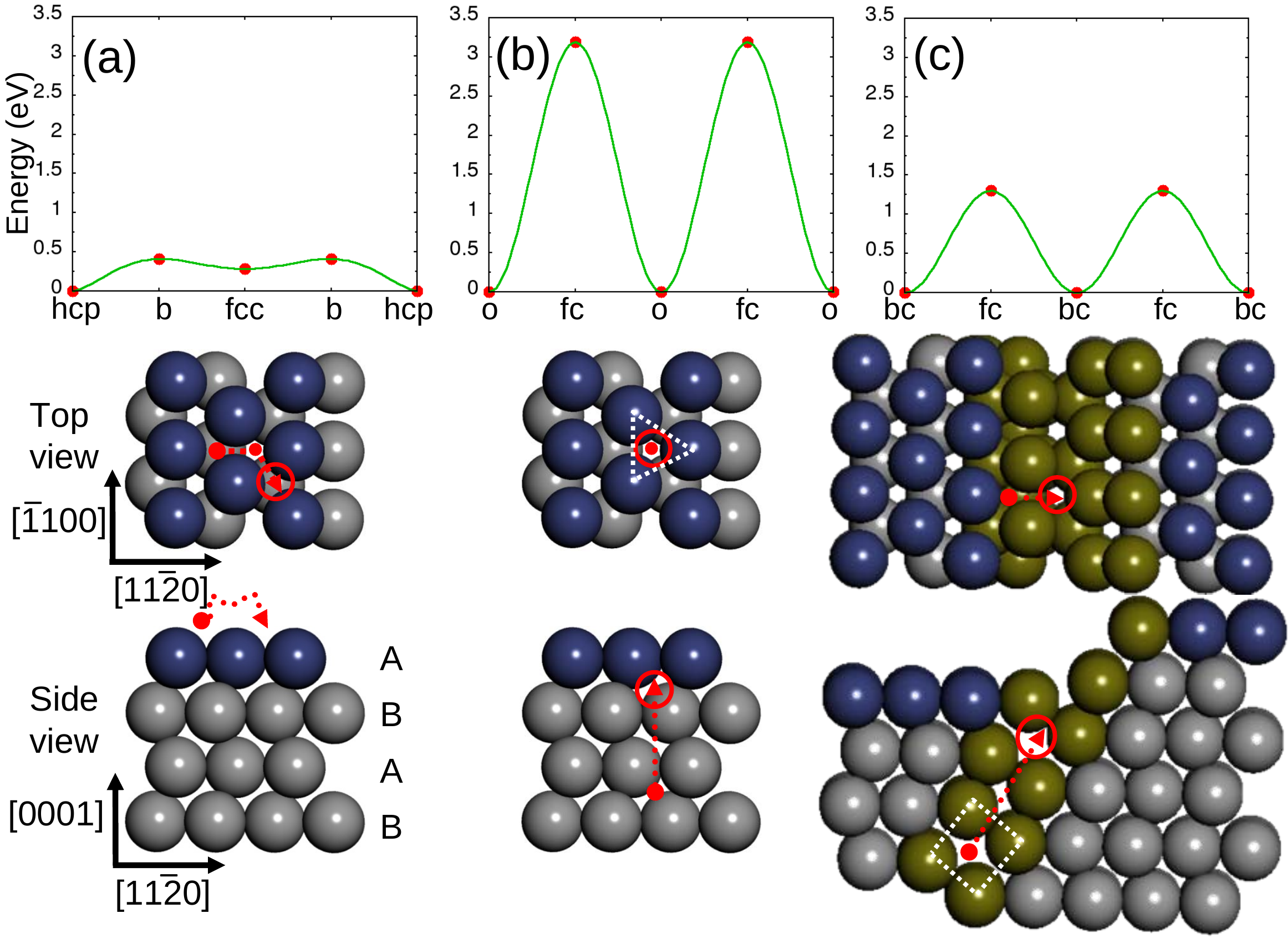}
\caption{(Color online) Surface and bulk diffusion of C atoms on a
carbon saturated Co(0001) surface. Surface diffusion in (a) is
compared to bulk diffusion in (b) and diffusion along a grain
boundary in (c). The top panels represent energy changes per atom
along the optimum diffusion path, which is indicated by the dashed
line in top and side views, presented in the bottom two panels.
\label{fig3}}
\end{figure}
%===========< FIGURE 3 >=================================================

Since a realistic representation of the growth mechanism by
molecular dynamics (MD) simulations is currently not possible due
to time limitations, we discuss in the following likely processes
that should contribute to foam growth and judge their importance
according to potential energy surfaces. To favor foam growth, we
need to find a suitable substrate geometry and identify growth
conditions that promote the formation of foam rather than other
competing
nanostructures\cite{{Banhart2011},{Banhart2007},{Ding2011}}.
Assuming that the feedstock are carbon atoms dissolved in the
substrate, we consider grain boundaries and steps as preferential
nucleation sites of the foam. Three competing processes contribute
to the nucleation and growth of carbon nanostructures on the
surface: surface diffusion of carbon, bulk diffusion of carbon
inside individual grains, and bulk diffusion along grain
boundaries that had not been considered previously.

Our results for these three processes are presented in
Fig.~\ref{fig3}. Since surface diffusion of C atoms, depicted in
Fig.~\ref{fig3}(a), does not require displacement of substrate
metal atoms, it occurs with a low activation barrier of only
$0.41$~eV and should be the fastest process of all. The optimum
path involves diffusion between the more stable hollow sites, with
the {\em hcp} sites being energetically favored by $0.28$~eV over
the {\em fcc} sites, across less stable bridge sites labeled {\em
b}.

In bulk cobalt, carbon atoms prefer energetically the octahedral
interstitial sites over the tetrahedral sites. The optimum bulk
path, presented in Fig.~\ref{fig3}(b), involves diffusion normal
to the surface between octahedral ({\em o}) sites across barriers
at the triangular face centers ({\em fc}) of the octahedra. We
emphasized one triangular face of an octahedron by the white
dotted line in the middle panel of Fig.~\ref{fig3}(b). In this
view, the barrier {\em fc} site in the center of the triangle
separates the favored {\em o} sites directly below and above.
Since the Co atoms are closely packed in the hcp structure,
passing through the center of the triangular face requires
displacing atoms, which requires a high activation energy of
$3.19$~eV. This value is to be considered an upper limit, since
presence of defects including vacancies should reduce the
activation barrier for bulk diffusion
significantly\cite{PCCP-Li2010}.

In contrast to a single crystal, the atomic packing at grain
boundaries is less compact. Consequently, interstitial carbon
atoms may find an energetically less costly diffusion path along
the grain boundary than in the perfect lattice. A possible grain
boundary structure that ends in a step edge is shown in the middle
and bottom panel of Fig.~\ref{fig3}(c). The atomic packing in this
grain boundary resembles that of a simple cubic lattice, with
interstitial carbon favoring energetically the body center {\em
bc} sites in the cube center. The optimum diffusion path requires
passing through a square face center {\em fc} at the interface of
neighboring cubes. As seen in the top panel of Fig.~\ref{fig3}(c),
the activation barrier for the diffusion along this grain boundary
is ${\approx}1.3$~eV, less than half the single crystal value.
Considering growth conditions similar to those in
Ref.~\onlinecite{Banhart2011},
% Temperature above 600 degC = 900 K
diffusion to the surface along this grain boundary should be
${\approx}4{\times}10^{10}$ times faster than in the perfect
crystal at $T=900$~K according to Arrhenius law.

%===========< FIGURE 4 >=================================================
% entire page. There is no need to do explicit centering.
\begin{figure}[tb]
\includegraphics[width=1.0\columnwidth]{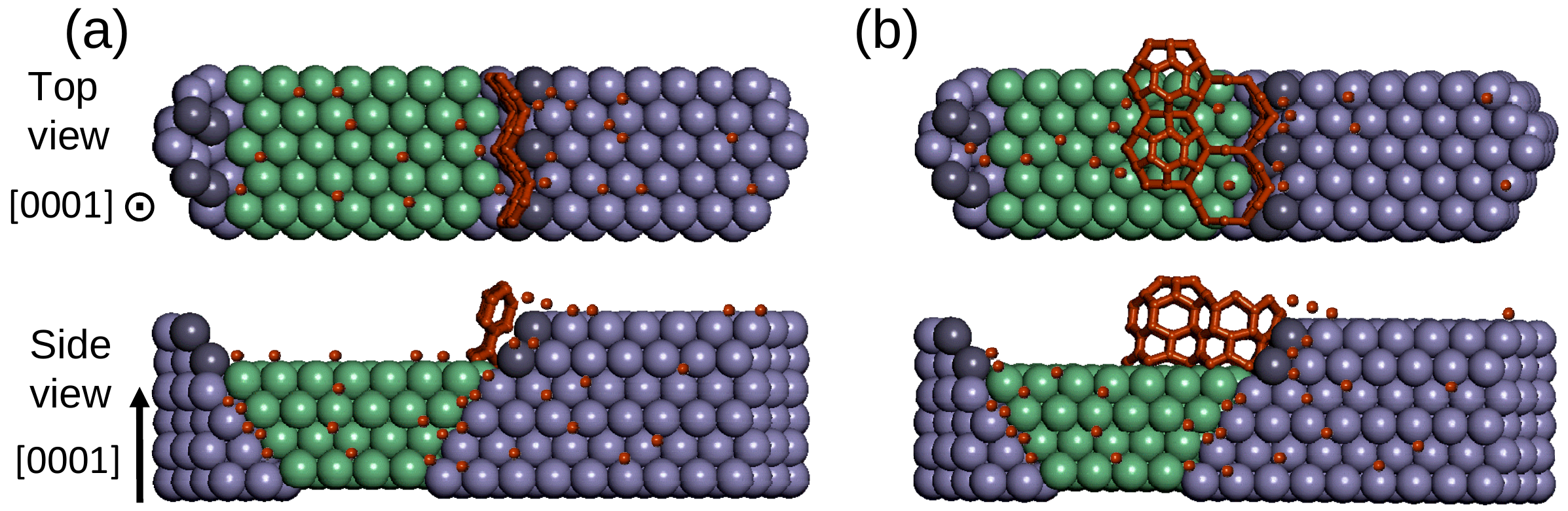}
\caption{(Color online) Possible formation mechanism of the
cellular carbon nanofoam, represented by structural snap shots in
top and side view. Different grains are distinguished by color and
shading. Initial formation of a graphene nanoribbon along a step
edge in (a) is followed by lateral growth of honeycomb cells in
(b). \label{fig4}}
\end{figure}
%===========< FIGURE 4 >=================================================

With the information at hand about the diffusion rates of the
carbon feedstock, we proceed to discuss a possible growth
scenario. The Co structure in Fig.~\ref{fig4} schematically
depicts three grains, distinguished by color and shading. It is
plausible to assume that the terrace height at both sides of the
grain boundary may not be the same, yielding a step structure,
which is best visible in side view. Under growth
conditions\cite{Banhart2011} near $600^\circ$C, the fastest rate
of carbon diffusion to the surface is along the grain boundary
towards the step edge, where carbon may aggregate to a narrow
graphene nanoribbon. Since according to our studies a zigzag edge
binds more strongly to Co than an armchair edge, we consider a
zigzag graphene nanoribbon attached to the step edge, as seen in
Fig.~\ref{fig4}(a). To best conform to the substrate, the
nanoribbon acquires a washboard structure, depicted in the top
panel in Fig.~\ref{fig4}(a). The more reactive nanoribbon atoms,
which protrude towards the terrace, are more likely to form bonds
with carbon atoms diffusing along the terrace, thus initiating the
formation of foam cells. In the meantime, atoms or flakes
diffusing along the upper terrace become the feedstock for the
termination of the foam layer by caps, as seen in
Fig.~\ref{fig4}(b). More detailed structure information and an
animation of the growth process is presented in the EPAPS on-line
material\cite{EPAPS-foamgrowth12}. We hope that this information
may encourage follow-up experimental studies aiming at
synthesizing the carbon foam and related carbon allotropes.

In conclusion, we studied the formation and structural as well as
thermal stability of cellular foam-like carbon nanostructures by
performing {\em ab initio} density functional calculations. We
found that these systems with a mixed $sp^2$/$sp^3$ bonding
character may be compressed by reducing the symmetry of the
honeycomb cells. The foam may accommodate the same type of defects
as graphene, and its surface may be be stabilized by terminating
caps. We postulate that the foam may form under non-equilibrium
conditions near grain boundaries on a carbon-saturated Co surface
and should be thermally stable up to ${\approx}3,700$~K.

\begin{acknowledgments}
We acknowledge extensive discussions with Florian Banhart, Julio
A. Rodr\'{\i}guez-Manzo, and Arkady V. Krasheninnikov.
% Fig.~\ref{fig4}(c) was provided by Florian Banhart and
% Julio A. Rodr\'{\i}guez-Manzo.
This work was supported by the National Science Foundation
Cooperative Agreement \#EEC-0832785, titled ``NSEC: Center for
High-rate Nanomanufacturing''. Computational resources have been
provided by the Michigan State University High Performance
Computing Center.
\end{acknowledgments}

%+++++++++++++++++++++++++++++++++++++++++++++++++++++++++++++++++++++
% You should use BibTeX and apsrev.bst for references
% Choosing a journal automatically selects the correct APS
% BibTeX style file (bst file), so only uncomment the line
% below if necessary.
% \bibliographystyle{apsrev4-1}
% \bibliography{foamgrowth12} %your bib file here

\begin{thebibliography}{32}%
\makeatletter
\providecommand \@ifxundefined [1]{%
 \@ifx{#1\undefined}
}%
\providecommand \@ifnum [1]{%
 \ifnum #1\expandafter \@firstoftwo
 \else \expandafter \@secondoftwo
 \fi
}%
\providecommand \@ifx [1]{%
 \ifx #1\expandafter \@firstoftwo
 \else \expandafter \@secondoftwo
 \fi
}%
\providecommand \natexlab [1]{#1}%
\providecommand \enquote  [1]{``#1''}%
\providecommand \bibnamefont  [1]{#1}%
\providecommand \bibfnamefont [1]{#1}%
\providecommand \citenamefont [1]{#1}%
\providecommand \href@noop [0]{\@secondoftwo}%
\providecommand \href [0]{\begingroup \@sanitize@url \@href}%
\providecommand \@href[1]{\@@startlink{#1}\@@href}%
\providecommand \@@href[1]{\endgroup#1\@@endlink}%
\providecommand \@sanitize@url [0]{\catcode `\\12\catcode
`\$12\catcode
  `\&12\catcode `\#12\catcode `\^12\catcode `\_12\catcode `\%12\relax}%
\providecommand \@@startlink[1]{}%
\providecommand \@@endlink[0]{}%
\providecommand \url  [0]{\begingroup\@sanitize@url \@url }%
\providecommand \@url [1]{\endgroup\@href {#1}{\urlprefix }}%
\providecommand \urlprefix  [0]{URL }%
\providecommand \Eprint [0]{\href }%
\providecommand \doibase [0]{http://dx.doi.org/}%
\providecommand \selectlanguage [0]{\@gobble}%
\providecommand \bibinfo  [0]{\@secondoftwo}%
\providecommand \bibfield  [0]{\@secondoftwo}%
\providecommand \translation [1]{[#1]}%
\providecommand \BibitemOpen [0]{}%
\providecommand \bibitemStop [0]{}%
\providecommand \bibitemNoStop [0]{.\EOS\space}%
\providecommand \EOS [0]{\spacefactor3000\relax}%
\providecommand \BibitemShut  [1]{\csname bibitem#1\endcsname}%
\let\auto@bib@innerbib\@empty
%</preamble>
\bibitem [{\citenamefont {Kroto}\ \emph {et~al.}(1985)\citenamefont {Kroto},
  \citenamefont {Heath}, \citenamefont {O'Brien}, \citenamefont {Curl},\ and\
  \citenamefont {Smalley}}]{Kroto85}%
  \BibitemOpen
  \bibfield  {author} {\bibinfo {author} {\bibfnamefont {H.~W.}\ \bibnamefont
  {Kroto}}, \bibinfo {author} {\bibfnamefont {J.~R.}\ \bibnamefont {Heath}},
  \bibinfo {author} {\bibfnamefont {S.~C.}\ \bibnamefont {O'Brien}}, \bibinfo
  {author} {\bibfnamefont {R.~F.}\ \bibnamefont {Curl}}, \ and\ \bibinfo
  {author} {\bibfnamefont {R.~E.}\ \bibnamefont {Smalley}},\ }\href {\doibase
  10.1038/318162a0} {\bibfield  {journal} {\bibinfo  {journal} {Nature}\
  }\textbf {\bibinfo {volume} {318}},\ \bibinfo {pages} {162} (\bibinfo {year}
  {1985})}\BibitemShut {NoStop}%
\bibitem [{\citenamefont {Iijima}(1991)}]{Iijima91}%
  \BibitemOpen
  \bibfield  {author} {\bibinfo {author} {\bibfnamefont {S.}~\bibnamefont
  {Iijima}},\ }\href {\doibase 10.1038/354056a0} {\bibfield  {journal}
  {\bibinfo  {journal} {Nature}\ }\textbf {\bibinfo {volume} {354}},\ \bibinfo
  {pages} {56} (\bibinfo {year} {1991})}\BibitemShut {NoStop}%
\bibitem [{\citenamefont {Novoselov}\ \emph {et~al.}(2004)\citenamefont
  {Novoselov}, \citenamefont {Geim}, \citenamefont {Morozov}, \citenamefont
  {Jiang}, \citenamefont {Zhang}, \citenamefont {Dubonos}, \citenamefont
  {Grigorieva},\ and\ \citenamefont {Firsove}}]{Geim04}%
  \BibitemOpen
  \bibfield  {author} {\bibinfo {author} {\bibfnamefont {K.~S.}\ \bibnamefont
  {Novoselov}}, \bibinfo {author} {\bibfnamefont {A.~K.}\ \bibnamefont {Geim}},
  \bibinfo {author} {\bibfnamefont {S.~V.}\ \bibnamefont {Morozov}}, \bibinfo
  {author} {\bibfnamefont {D.}~\bibnamefont {Jiang}}, \bibinfo {author}
  {\bibfnamefont {Y.}~\bibnamefont {Zhang}}, \bibinfo {author} {\bibfnamefont
  {S.~V.}\ \bibnamefont {Dubonos}}, \bibinfo {author} {\bibfnamefont {I.~V.}\
  \bibnamefont {Grigorieva}}, \ and\ \bibinfo {author} {\bibfnamefont {A.~A.}\
  \bibnamefont {Firsove}},\ }\href {\doibase 10.1126/science.1102896}
  {\bibfield  {journal} {\bibinfo  {journal} {Science}\ }\textbf {\bibinfo
  {volume} {306}},\ \bibinfo {pages} {666} (\bibinfo {year}
  {2004})}\BibitemShut {NoStop}%
\bibitem [{\citenamefont {Umemoto}\ \emph {et~al.}(2001)\citenamefont
  {Umemoto}, \citenamefont {Saito}, \citenamefont {Berber},\ and\ \citenamefont
  {Tomanek}}]{DT141}%
  \BibitemOpen
  \bibfield  {author} {\bibinfo {author} {\bibfnamefont {K.}~\bibnamefont
  {Umemoto}}, \bibinfo {author} {\bibfnamefont {S.}~\bibnamefont {Saito}},
  \bibinfo {author} {\bibfnamefont {S.}~\bibnamefont {Berber}}, \ and\ \bibinfo
  {author} {\bibfnamefont {D.}~\bibnamefont {Tomanek}},\ }\href {\doibase
  10.1103/PhysRevB.64.193409} {\bibfield  {journal} {\bibinfo  {journal} {Phys.
  Rev. B}\ }\textbf {\bibinfo {volume} {64}},\ \bibinfo {pages} {193409}
  (\bibinfo {year} {2001})}\BibitemShut {NoStop}%
\bibitem [{\citenamefont {Kuc}\ and\ \citenamefont
  {Seifert}(2006)}]{Seifert2006}%
  \BibitemOpen
  \bibfield  {author} {\bibinfo {author} {\bibfnamefont {A.}~\bibnamefont
  {Kuc}}\ and\ \bibinfo {author} {\bibfnamefont {G.}~\bibnamefont {Seifert}},\
  }\href {\doibase 10.1103/PhysRevB.74.214104} {\bibfield  {journal} {\bibinfo
  {journal} {Phys. Rev. B}\ }\textbf {\bibinfo {volume} {74}},\ \bibinfo
  {pages} {214104} (\bibinfo {year} {2006})}\BibitemShut {NoStop}%
\bibitem [{\citenamefont {Zhao}\ \emph {et~al.}(2011)\citenamefont {Zhao},
  \citenamefont {Xu}, \citenamefont {Wang}, \citenamefont {Zhou}, \citenamefont
  {He}, \citenamefont {Liu}, \citenamefont {Wang},\ and\ \citenamefont
  {Tian}}]{Zhao2011}%
  \BibitemOpen
  \bibfield  {author} {\bibinfo {author} {\bibfnamefont {Z.}~\bibnamefont
  {Zhao}}, \bibinfo {author} {\bibfnamefont {B.}~\bibnamefont {Xu}}, \bibinfo
  {author} {\bibfnamefont {L.-M.}\ \bibnamefont {Wang}}, \bibinfo {author}
  {\bibfnamefont {X.-F.}\ \bibnamefont {Zhou}}, \bibinfo {author}
  {\bibfnamefont {J.}~\bibnamefont {He}}, \bibinfo {author} {\bibfnamefont
  {Z.}~\bibnamefont {Liu}}, \bibinfo {author} {\bibfnamefont {H.-T.}\
  \bibnamefont {Wang}}, \ and\ \bibinfo {author} {\bibfnamefont
  {Y.}~\bibnamefont {Tian}},\ }\href {\doibase 10.1021/nn202053t} {\bibfield
  {journal} {\bibinfo  {journal} {ACS Nano}\ }\textbf {\bibinfo {volume} {5}},\
  \bibinfo {pages} {7226} (\bibinfo {year} {2011})}\BibitemShut {NoStop}%
\bibitem [{\citenamefont {Enyashin}\ and\ \citenamefont
  {Ivanovskii}(2011)}]{Ivanovskii2011}%
  \BibitemOpen
  \bibfield  {author} {\bibinfo {author} {\bibfnamefont {A.~N.}\ \bibnamefont
  {Enyashin}}\ and\ \bibinfo {author} {\bibfnamefont {A.~L.}\ \bibnamefont
  {Ivanovskii}},\ }\href {\doibase 10.1002/pssb.201046583} {\bibfield
  {journal} {\bibinfo  {journal} {Phys. Stat. Sol. (b)}\ }\textbf {\bibinfo
  {volume} {248}},\ \bibinfo {pages} {1879} (\bibinfo {year}
  {2011})}\BibitemShut {NoStop}%
\bibitem [{\citenamefont {Rodr\'{\i}guez-Manzo}\ \emph
  {et~al.}(2011)\citenamefont {Rodr\'{\i}guez-Manzo}, \citenamefont
  {Pham-Huu},\ and\ \citenamefont {Banhart}}]{Banhart2011}%
  \BibitemOpen
  \bibfield  {author} {\bibinfo {author} {\bibfnamefont {J.~A.}\ \bibnamefont
  {Rodr\'{\i}guez-Manzo}}, \bibinfo {author} {\bibfnamefont {C.}~\bibnamefont
  {Pham-Huu}}, \ and\ \bibinfo {author} {\bibfnamefont {F.}~\bibnamefont
  {Banhart}},\ }\href {\doibase 10.1021/nn103456z} {\bibfield  {journal}
  {\bibinfo  {journal} {ACS Nano}\ }\textbf {\bibinfo {volume} {5}},\ \bibinfo
  {pages} {1529} (\bibinfo {year} {2011})}\BibitemShut {NoStop}%
\bibitem [{\citenamefont {Rodriguez-Manzo}\ \emph {et~al.}(2007)\citenamefont
  {Rodriguez-Manzo}, \citenamefont {Terrones}, \citenamefont {Terrones},
  \citenamefont {Kroto}, \citenamefont {Sun},\ and\ \citenamefont
  {Banhart}}]{Banhart2007}%
  \BibitemOpen
  \bibfield  {author} {\bibinfo {author} {\bibfnamefont {J.~A.}\ \bibnamefont
  {Rodriguez-Manzo}}, \bibinfo {author} {\bibfnamefont {M.}~\bibnamefont
  {Terrones}}, \bibinfo {author} {\bibfnamefont {H.}~\bibnamefont {Terrones}},
  \bibinfo {author} {\bibfnamefont {H.~W.}\ \bibnamefont {Kroto}}, \bibinfo
  {author} {\bibfnamefont {L.}~\bibnamefont {Sun}}, \ and\ \bibinfo {author}
  {\bibfnamefont {F.}~\bibnamefont {Banhart}},\ }\href {\doibase
  10.1038/nnano.2007.107} {\bibfield  {journal} {\bibinfo  {journal} {Nature
  Nano.}\ }\textbf {\bibinfo {volume} {2}},\ \bibinfo {pages} {307} (\bibinfo
  {year} {2007})}\BibitemShut {NoStop}%
\bibitem [{\citenamefont {Lin}\ \emph {et~al.}(2007)\citenamefont {Lin},
  \citenamefont {Tan}, \citenamefont {Boothroyd}, \citenamefont {Loh},
  \citenamefont {Tok},\ and\ \citenamefont {Foo}}]{LinNL2007}%
  \BibitemOpen
  \bibfield  {author} {\bibinfo {author} {\bibfnamefont {M.}~\bibnamefont
  {Lin}}, \bibinfo {author} {\bibfnamefont {J.~P.~Y.}\ \bibnamefont {Tan}},
  \bibinfo {author} {\bibfnamefont {C.}~\bibnamefont {Boothroyd}}, \bibinfo
  {author} {\bibfnamefont {K.~P.}\ \bibnamefont {Loh}}, \bibinfo {author}
  {\bibfnamefont {E.~S.}\ \bibnamefont {Tok}}, \ and\ \bibinfo {author}
  {\bibfnamefont {Y.-L.}\ \bibnamefont {Foo}},\ }\href {\doibase
  10.1021/nl070681x} {\bibfield  {journal} {\bibinfo  {journal} {Nano Lett.}\
  }\textbf {\bibinfo {volume} {7}},\ \bibinfo {pages} {2234} (\bibinfo {year}
  {2007})}\BibitemShut {NoStop}%
\bibitem [{\citenamefont {Helveg}\ \emph {et~al.}(2004)\citenamefont {Helveg},
  \citenamefont {Lopez-Cartes}, \citenamefont {Sehested}, \citenamefont
  {Hansen}, \citenamefont {Clausen}, \citenamefont {Rostrup-Nielsen},
  \citenamefont {Abild-Pedersen},\ and\ \citenamefont {Norskov}}]{Helveg2004}%
  \BibitemOpen
  \bibfield  {author} {\bibinfo {author} {\bibfnamefont {S.}~\bibnamefont
  {Helveg}}, \bibinfo {author} {\bibfnamefont {C.}~\bibnamefont
  {Lopez-Cartes}}, \bibinfo {author} {\bibfnamefont {J.}~\bibnamefont
  {Sehested}}, \bibinfo {author} {\bibfnamefont {P.}~\bibnamefont {Hansen}},
  \bibinfo {author} {\bibfnamefont {B.}~\bibnamefont {Clausen}}, \bibinfo
  {author} {\bibfnamefont {J.}~\bibnamefont {Rostrup-Nielsen}}, \bibinfo
  {author} {\bibfnamefont {F.}~\bibnamefont {Abild-Pedersen}}, \ and\ \bibinfo
  {author} {\bibfnamefont {J.}~\bibnamefont {Norskov}},\ }\href {\doibase
  10.1038/Nature02278} {\bibfield  {journal} {\bibinfo  {journal} {Nature}\
  }\textbf {\bibinfo {volume} {427}},\ \bibinfo {pages} {426} (\bibinfo {year}
  {2004})}\BibitemShut {NoStop}%
\bibitem [{\citenamefont {Artacho}\ \emph {et~al.}(2008)\citenamefont
  {Artacho}, \citenamefont {Anglada}, \citenamefont {Dieguez}, \citenamefont
  {Gale}, \citenamefont {Garcia}, \citenamefont {Junquera}, \citenamefont
  {Martin}, \citenamefont {Ordejon}, \citenamefont {Pruneda}, \citenamefont
  {Sanchez-Portal},\ and\ \citenamefont {Soler}}]{SIESTA}%
  \BibitemOpen
  \bibfield  {author} {\bibinfo {author} {\bibfnamefont {E.}~\bibnamefont
  {Artacho}}, \bibinfo {author} {\bibfnamefont {E.}~\bibnamefont {Anglada}},
  \bibinfo {author} {\bibfnamefont {O.}~\bibnamefont {Dieguez}}, \bibinfo
  {author} {\bibfnamefont {J.~D.}\ \bibnamefont {Gale}}, \bibinfo {author}
  {\bibfnamefont {A.}~\bibnamefont {Garcia}}, \bibinfo {author} {\bibfnamefont
  {J.}~\bibnamefont {Junquera}}, \bibinfo {author} {\bibfnamefont {R.~M.}\
  \bibnamefont {Martin}}, \bibinfo {author} {\bibfnamefont {P.}~\bibnamefont
  {Ordejon}}, \bibinfo {author} {\bibfnamefont {J.~M.}\ \bibnamefont
  {Pruneda}}, \bibinfo {author} {\bibfnamefont {D.}~\bibnamefont
  {Sanchez-Portal}}, \ and\ \bibinfo {author} {\bibfnamefont {J.~M.}\
  \bibnamefont {Soler}},\ }\href {\doibase 10.1088/0953-8984/20/6/064208}
  {\bibfield  {journal} {\bibinfo  {journal} {J. Phys. Cond. Mat.}\ }\textbf
  {\bibinfo {volume} {20}},\ \bibinfo {pages} {064208} (\bibinfo {year}
  {2008})}\BibitemShut {NoStop}%
\bibitem [{\citenamefont {Ceperley}\ and\ \citenamefont
  {Alder}(1980)}]{Ceperley1980}%
  \BibitemOpen
  \bibfield  {author} {\bibinfo {author} {\bibfnamefont {D.~M.}\ \bibnamefont
  {Ceperley}}\ and\ \bibinfo {author} {\bibfnamefont {B.~J.}\ \bibnamefont
  {Alder}},\ }\href {\doibase 10.1103/PhysRevLett.45.566} {\bibfield  {journal}
  {\bibinfo  {journal} {Phys. Rev. Lett.}\ }\textbf {\bibinfo {volume} {45}},\
  \bibinfo {pages} {566} (\bibinfo {year} {1980})}\BibitemShut {NoStop}%
\bibitem [{\citenamefont {Perdew}\ and\ \citenamefont
  {Zunger}(1981)}]{Perdew81}%
  \BibitemOpen
  \bibfield  {author} {\bibinfo {author} {\bibfnamefont {J.~P.}\ \bibnamefont
  {Perdew}}\ and\ \bibinfo {author} {\bibfnamefont {A.}~\bibnamefont
  {Zunger}},\ }\href@noop {} {\bibfield  {journal} {\bibinfo  {journal} {Phys.
  Rev. B}\ }\textbf {\bibinfo {volume} {23}},\ \bibinfo {pages} {5048}
  (\bibinfo {year} {1981})}\BibitemShut {NoStop}%
\bibitem [{\citenamefont {Troullier}\ and\ \citenamefont
  {Martins}(1991)}]{Troullier91}%
  \BibitemOpen
  \bibfield  {author} {\bibinfo {author} {\bibfnamefont {N.}~\bibnamefont
  {Troullier}}\ and\ \bibinfo {author} {\bibfnamefont {J.~L.}\ \bibnamefont
  {Martins}},\ }\href@noop {} {\bibfield  {journal} {\bibinfo  {journal} {Phys.
  Rev. B}\ }\textbf {\bibinfo {volume} {43}},\ \bibinfo {pages} {1993}
  (\bibinfo {year} {1991})}\BibitemShut {NoStop}%
\bibitem [{\citenamefont {Monkhorst}\ and\ \citenamefont
  {Pack}(1976)}]{Monkhorst-Pack76}%
  \BibitemOpen
  \bibfield  {author} {\bibinfo {author} {\bibfnamefont {H.~J.}\ \bibnamefont
  {Monkhorst}}\ and\ \bibinfo {author} {\bibfnamefont {J.~D.}\ \bibnamefont
  {Pack}},\ }\href {\doibase 10.1103/PhysRevB.13.5188} {\bibfield  {journal}
  {\bibinfo  {journal} {Phys. Rev. B}\ }\textbf {\bibinfo {volume} {13}},\
  \bibinfo {pages} {5188} (\bibinfo {year} {1976})}\BibitemShut {NoStop}%
\bibitem [{\citenamefont {Kittel}(2005)}]{Kittel05}%
  \BibitemOpen
  \bibfield  {author} {\bibinfo {author} {\bibfnamefont {C.}~\bibnamefont
  {Kittel}},\ }\href@noop {} {\emph {\bibinfo {title} {Introduction to Solid
  State Physics}}},\ \bibinfo {edition} {eighth}\ ed.\ (\bibinfo  {publisher}
  {Wiley},\ \bibinfo {address} {New York},\ \bibinfo {year} {2005})\BibitemShut
  {NoStop}%
\bibitem [{\citenamefont {Okada}\ and\ \citenamefont
  {Oshiyama}(2003)}]{Okada03}%
  \BibitemOpen
  \bibfield  {author} {\bibinfo {author} {\bibfnamefont {S.}~\bibnamefont
  {Okada}}\ and\ \bibinfo {author} {\bibfnamefont {A.}~\bibnamefont
  {Oshiyama}},\ }\href {\doibase 10.1143/JPSJ.72.1510} {\bibfield  {journal}
  {\bibinfo  {journal} {J. Phys. Soc. Jpn.}\ }\textbf {\bibinfo {volume}
  {72}},\ \bibinfo {pages} {1510} (\bibinfo {year} {2003})}\BibitemShut
  {NoStop}%
\bibitem [{\citenamefont {Higuchi}\ \emph {et~al.}(2004)\citenamefont
  {Higuchi}, \citenamefont {Kusakabe}, \citenamefont {Suzuki}, \citenamefont
  {Tsuneyuki}, \citenamefont {Yamauchi}, \citenamefont {Akagi},\ and\
  \citenamefont {Yoshimoto}}]{Higuchi04}%
  \BibitemOpen
  \bibfield  {author} {\bibinfo {author} {\bibfnamefont {Y.}~\bibnamefont
  {Higuchi}}, \bibinfo {author} {\bibfnamefont {K.}~\bibnamefont {Kusakabe}},
  \bibinfo {author} {\bibfnamefont {N.}~\bibnamefont {Suzuki}}, \bibinfo
  {author} {\bibfnamefont {S.}~\bibnamefont {Tsuneyuki}}, \bibinfo {author}
  {\bibfnamefont {J.}~\bibnamefont {Yamauchi}}, \bibinfo {author}
  {\bibfnamefont {K.}~\bibnamefont {Akagi}}, \ and\ \bibinfo {author}
  {\bibfnamefont {Y.}~\bibnamefont {Yoshimoto}},\ }\href {\doibase
  10.1088/0953-8984/16/48/028} {\bibfield  {journal} {\bibinfo  {journal} {J.
  Phys. Cond. Mat.}\ }\textbf {\bibinfo {volume} {16}},\ \bibinfo {pages}
  {S5689} (\bibinfo {year} {2004})}\BibitemShut {NoStop}%
\bibitem [{\citenamefont {Ma\~nanes}\ \emph {et~al.}(2008)\citenamefont
  {Ma\~nanes}, \citenamefont {Duque}, \citenamefont {Ayuela}, \citenamefont
  {L\'opez},\ and\ \citenamefont {Alonso}}]{AlonsoZZ08}%
  \BibitemOpen
  \bibfield  {author} {\bibinfo {author} {\bibfnamefont {A.}~\bibnamefont
  {Ma\~nanes}}, \bibinfo {author} {\bibfnamefont {F.}~\bibnamefont {Duque}},
  \bibinfo {author} {\bibfnamefont {A.}~\bibnamefont {Ayuela}}, \bibinfo
  {author} {\bibfnamefont {M.~J.}\ \bibnamefont {L\'opez}}, \ and\ \bibinfo
  {author} {\bibfnamefont {J.~A.}\ \bibnamefont {Alonso}},\ }\href {\doibase
  10.1103/PhysRevB.78.035432} {\bibfield  {journal} {\bibinfo  {journal} {Phys.
  Rev. B}\ }\textbf {\bibinfo {volume} {78}},\ \bibinfo {pages} {035432}
  (\bibinfo {year} {2008})}\BibitemShut {NoStop}%
\bibitem [{\citenamefont {Hod}\ and\ \citenamefont
  {Scuseria}(2008)}]{HodACSNN08}%
  \BibitemOpen
  \bibfield  {author} {\bibinfo {author} {\bibfnamefont {O.}~\bibnamefont
  {Hod}}\ and\ \bibinfo {author} {\bibfnamefont {G.~E.}\ \bibnamefont
  {Scuseria}},\ }\href {\doibase 10.1021/nn8004069} {\bibfield  {journal}
  {\bibinfo  {journal} {ACS Nano}\ }\textbf {\bibinfo {volume} {2}},\ \bibinfo
  {pages} {2243} (\bibinfo {year} {2008})}\BibitemShut {NoStop}%
\bibitem [{\citenamefont {Kim}\ \emph {et~al.}(2003)\citenamefont {Kim},
  \citenamefont {Choi}, \citenamefont {Chang},\ and\ \citenamefont
  {Tomanek}}]{DT160}%
  \BibitemOpen
  \bibfield  {author} {\bibinfo {author} {\bibfnamefont {Y.-H.}\ \bibnamefont
  {Kim}}, \bibinfo {author} {\bibfnamefont {J.}~\bibnamefont {Choi}}, \bibinfo
  {author} {\bibfnamefont {K.~J.}\ \bibnamefont {Chang}}, \ and\ \bibinfo
  {author} {\bibfnamefont {D.}~\bibnamefont {Tomanek}},\ }\href {\doibase
  10.1103/PhysRevB.68.125420} {\bibfield  {journal} {\bibinfo  {journal} {Phys.
  Rev. B}\ }\textbf {\bibinfo {volume} {68}},\ \bibinfo {pages} {125420}
  (\bibinfo {year} {2003})}\BibitemShut {NoStop}%
\bibitem [{\citenamefont {Stone}\ and\ \citenamefont {Wales}(1986)}]{Stone86}%
  \BibitemOpen
  \bibfield  {author} {\bibinfo {author} {\bibfnamefont {A.~J.}\ \bibnamefont
  {Stone}}\ and\ \bibinfo {author} {\bibfnamefont {D.~J.}\ \bibnamefont
  {Wales}},\ }\href@noop {} {\bibfield  {journal} {\bibinfo  {journal} {Chem.
  Phys. Lett.}\ }\textbf {\bibinfo {volume} {128}},\ \bibinfo {pages} {501}
  (\bibinfo {year} {1986})}\BibitemShut {NoStop}%
\bibitem [{\citenamefont {Thrower}(1969)}]{Thrower69}%
  \BibitemOpen
  \bibfield  {author} {\bibinfo {author} {\bibfnamefont {P.~A.}\ \bibnamefont
  {Thrower}},\ }in\ \href@noop {} {\emph {\bibinfo {booktitle} {Chemistry and
  physics of carbon, Vol. 5}}},\ \bibinfo {editor} {edited by\ \bibinfo
  {editor} {\bibfnamefont {P.~L.}\ \bibnamefont {Walker~Jr.}}}\ (\bibinfo
  {publisher} {Marcel Dekker},\ \bibinfo {address} {New York},\ \bibinfo {year}
  {1969})\ pp.\ \bibinfo {pages} {217--320}\BibitemShut {NoStop}%
\bibitem [{\citenamefont {Huang}\ \emph {et~al.}(2011)\citenamefont {Huang},
  \citenamefont {Ruiz-Vargas}, \citenamefont {van~der Zande}, \citenamefont
  {Whitney}, \citenamefont {Levendorf}, \citenamefont {Kevek}, \citenamefont
  {Garg}, \citenamefont {Alden}, \citenamefont {Hustedt}, \citenamefont {Zhu},
  \citenamefont {Park}, \citenamefont {McEuen},\ and\ \citenamefont
  {Muller}}]{Huang11}%
  \BibitemOpen
  \bibfield  {author} {\bibinfo {author} {\bibfnamefont {P.~Y.}\ \bibnamefont
  {Huang}}, \bibinfo {author} {\bibfnamefont {C.~S.}\ \bibnamefont
  {Ruiz-Vargas}}, \bibinfo {author} {\bibfnamefont {A.~M.}\ \bibnamefont
  {van~der Zande}}, \bibinfo {author} {\bibfnamefont {W.~S.}\ \bibnamefont
  {Whitney}}, \bibinfo {author} {\bibfnamefont {M.~P.}\ \bibnamefont
  {Levendorf}}, \bibinfo {author} {\bibfnamefont {J.~W.}\ \bibnamefont
  {Kevek}}, \bibinfo {author} {\bibfnamefont {S.}~\bibnamefont {Garg}},
  \bibinfo {author} {\bibfnamefont {J.~S.}\ \bibnamefont {Alden}}, \bibinfo
  {author} {\bibfnamefont {C.~J.}\ \bibnamefont {Hustedt}}, \bibinfo {author}
  {\bibfnamefont {Y.}~\bibnamefont {Zhu}}, \bibinfo {author} {\bibfnamefont
  {J.}~\bibnamefont {Park}}, \bibinfo {author} {\bibfnamefont {P.~L.}\
  \bibnamefont {McEuen}}, \ and\ \bibinfo {author} {\bibfnamefont {D.~A.}\
  \bibnamefont {Muller}},\ }\href {\doibase 10.1038/Nature09718} {\bibfield
  {journal} {\bibinfo  {journal} {Nature}\ }\textbf {\bibinfo {volume} {469}},\
  \bibinfo {pages} {389} (\bibinfo {year} {2011})}\BibitemShut {NoStop}%
\bibitem [{\citenamefont {Banhart}\ \emph {et~al.}(2011)\citenamefont
  {Banhart}, \citenamefont {Kotakoski},\ and\ \citenamefont
  {Krasheninnikov}}]{Arkady11}%
  \BibitemOpen
  \bibfield  {author} {\bibinfo {author} {\bibfnamefont {F.}~\bibnamefont
  {Banhart}}, \bibinfo {author} {\bibfnamefont {J.}~\bibnamefont {Kotakoski}},
  \ and\ \bibinfo {author} {\bibfnamefont {A.~V.}\ \bibnamefont
  {Krasheninnikov}},\ }\href {\doibase 10.1021/nn102598m} {\bibfield  {journal}
  {\bibinfo  {journal} {ACS Nano}\ }\textbf {\bibinfo {volume} {5}},\ \bibinfo
  {pages} {26} (\bibinfo {year} {2011})}\BibitemShut {NoStop}%
\bibitem [{\citenamefont {Meyer}\ \emph {et~al.}(2008)\citenamefont {Meyer},
  \citenamefont {Kisielowski}, \citenamefont {Erni}, \citenamefont {Rossell},
  \citenamefont {Crommie},\ and\ \citenamefont {Zettl}}]{Meyer08}%
  \BibitemOpen
  \bibfield  {author} {\bibinfo {author} {\bibfnamefont {J.~C.}\ \bibnamefont
  {Meyer}}, \bibinfo {author} {\bibfnamefont {C.}~\bibnamefont {Kisielowski}},
  \bibinfo {author} {\bibfnamefont {R.}~\bibnamefont {Erni}}, \bibinfo {author}
  {\bibfnamefont {M.~D.}\ \bibnamefont {Rossell}}, \bibinfo {author}
  {\bibfnamefont {M.~F.}\ \bibnamefont {Crommie}}, \ and\ \bibinfo {author}
  {\bibfnamefont {A.}~\bibnamefont {Zettl}},\ }\href {\doibase
  10.1021/nl801386m} {\bibfield  {journal} {\bibinfo  {journal} {Nano Lett.}\
  }\textbf {\bibinfo {volume} {8}},\ \bibinfo {pages} {3582} (\bibinfo {year}
  {2008})}\BibitemShut {NoStop}%
\bibitem [{\citenamefont {Lahiri}\ \emph {et~al.}(2010)\citenamefont {Lahiri},
  \citenamefont {Lin}, \citenamefont {Bozkurt}, \citenamefont {Oleynik},\ and\
  \citenamefont {Batzill}}]{Lahiri10}%
  \BibitemOpen
  \bibfield  {author} {\bibinfo {author} {\bibfnamefont {J.}~\bibnamefont
  {Lahiri}}, \bibinfo {author} {\bibfnamefont {Y.}~\bibnamefont {Lin}},
  \bibinfo {author} {\bibfnamefont {P.}~\bibnamefont {Bozkurt}}, \bibinfo
  {author} {\bibfnamefont {I.~I.}\ \bibnamefont {Oleynik}}, \ and\ \bibinfo
  {author} {\bibfnamefont {M.}~\bibnamefont {Batzill}},\ }\href {\doibase
  10.1038/NNANO.2010.53} {\bibfield  {journal} {\bibinfo  {journal} {Nature
  Nano.}\ }\textbf {\bibinfo {volume} {5}},\ \bibinfo {pages} {326} (\bibinfo
  {year} {2010})}\BibitemShut {NoStop}%
\bibitem [{\citenamefont {Tersoff}(1988)}]{tersoff1988}%
  \BibitemOpen
  \bibfield  {author} {\bibinfo {author} {\bibfnamefont {J.}~\bibnamefont
  {Tersoff}},\ }\href {\doibase 10.1103/PhysRevLett.61.2879} {\bibfield
  {journal} {\bibinfo  {journal} {Phys. Rev. Lett.}\ }\textbf {\bibinfo
  {volume} {61}},\ \bibinfo {pages} {2879} (\bibinfo {year}
  {1988})}\BibitemShut {NoStop}%
\bibitem [{EPA()}]{EPAPS-foamgrowth12}%
  \BibitemOpen
  \href@noop {} {}\bibinfo {note} {See the supplementary information for
  molecluar dynamics simulations of the thermal stability of the foam and
  movies illustrating the growth mechanism.}\BibitemShut {Stop}%
\bibitem [{\citenamefont {Gao}\ \emph {et~al.}(2011)\citenamefont {Gao},
  \citenamefont {Yip}, \citenamefont {Zhao}, \citenamefont {Yakobson},\ and\
  \citenamefont {Ding}}]{Ding2011}%
  \BibitemOpen
  \bibfield  {author} {\bibinfo {author} {\bibfnamefont {J.}~\bibnamefont
  {Gao}}, \bibinfo {author} {\bibfnamefont {J.}~\bibnamefont {Yip}}, \bibinfo
  {author} {\bibfnamefont {J.}~\bibnamefont {Zhao}}, \bibinfo {author}
  {\bibfnamefont {B.~I.}\ \bibnamefont {Yakobson}}, \ and\ \bibinfo {author}
  {\bibfnamefont {F.}~\bibnamefont {Ding}},\ }\href {\doibase
  10.1021/ja110927p} {\bibfield  {journal} {\bibinfo  {journal} {J. Am. Chem.
  Soc.}\ }\textbf {\bibinfo {volume} {133}},\ \bibinfo {pages} {5009} (\bibinfo
  {year} {2011})}\BibitemShut {NoStop}%
\bibitem [{\citenamefont {Li}\ \emph {et~al.}(2010)\citenamefont {Li},
  \citenamefont {Zhang}, \citenamefont {Chen}, \citenamefont {Cui},\ and\
  \citenamefont {Pan}}]{PCCP-Li2010}%
  \BibitemOpen
  \bibfield  {author} {\bibinfo {author} {\bibfnamefont {B.}~\bibnamefont
  {Li}}, \bibinfo {author} {\bibfnamefont {Q.}~\bibnamefont {Zhang}}, \bibinfo
  {author} {\bibfnamefont {L.}~\bibnamefont {Chen}}, \bibinfo {author}
  {\bibfnamefont {P.}~\bibnamefont {Cui}}, \ and\ \bibinfo {author}
  {\bibfnamefont {X.}~\bibnamefont {Pan}},\ }\href {\doibase 10.1039/B925764K}
  {\bibfield  {journal} {\bibinfo  {journal} {Phys. Chem. Chem. Phys.}\
  }\textbf {\bibinfo {volume} {12}},\ \bibinfo {pages} {7848} (\bibinfo {year}
  {2010})}\BibitemShut {NoStop}%
\end{thebibliography}
% \end{document}
%+++++++++++++++++++++++++++++++++++++++++++++++++++++++++++++++++++++


\end{document}